# Status of the technologies for the production of the Cherenkov Telescope Array (CTA) mirrors

G. Pareschi[1], T. Armstrong[2], H. Baba[3], J. Bähr[4], A. Bonardi[5], G. Bonnoli[1], P. Brun[6], R. Canestrari[1], P. Chadwick[2], M. Chikawa[7], P.-H. Carton[6], V. de Souza[8], J. Dipold[8], M. Doro[9], D. Durand[6], M. Dyrda[10], A. Förster[11], M. Garczarczyk[4], E. Giro[12], J.-F. Glicenstein[6], Y. Hanabata[13], M. Hayashida[7], M. Hrabovski[14], C. Jeanney[6], M. Kagaya[3], H. Katagiri[3], L. Lessio[12], D. Mandat[14], M. Mariotti[9], C. Medina[15], J. Michałowski[10], P. Micolon[6], D. Nakajima[16], J. Niemiec[10], A. Nozato[13], M. Palatka[14], M. Pech[14], B. Peyaud[6], G. Pühlhofer[5], M. Rataj[17], G. Rodeghiero[12], G. Rojas[18], J. Rousselle[19], R. Sakonaka[13], P. Schovanek[14], K. Seweryn[17], C. Schultz[9], S. Shu[13], F. Stinzing[20], M. Stodulski[10], M. Teshima[11], P. Travniczek[14], C. van Eldik[20], V. Vassiliev[19], Ł. Wiśniewski[17], A. Wörnlein[20], T. Yoshida[3]
FOR THE CTA CONSORTIUM.

[1] INAF/Brera Astronomical Observatory, Milano/ Merate, Italy
[2] University of Durham, Durham, UK
[3] Ibaraki University, Japan
[4] Desy Zeuthen, Zeuthen, Germany
[5] IAAT, Universität Tübingen, Tübingen, Germany
[6] CEA, Irfu, Centre de Saclay, Gif-sur-Yvette, France
[7] ICRR University of Tokyo, Japan
[8] Instituto de Fsica de So Carlos, Universidade de So Paulo, Brasil
[9] University of Padova and INFN sezione di Padova, Padova, Italy
[10] Institute of Nuclear Physics, Polish Academy of Sciences, Krakow, Poland
[11] Max-Planck-Institut für Kernphysik, Heidelberg, Germany
[12] INAF - Osservatorio Astronomico di Padova, Padova, Italy
[13] Kinki University, Japan
[14] Institute of Physics of the Academy of Sciences of the Czech Republic, Prague, Czech Republic
[15] Instituto Argentino de Radioastronomia, CCT La Plata-CONICET, Argentina
[16] Max-Planck-Institut für Physik, Mu ̈nchen, Germany
[17] Space Research Center - Polish Academy of Science, Warsaw, Poland
[18] Universidade Federal de So Carlos
[19] University of California Los Angeles, California, USA
[20] ECAP, Universität Erlangen-Nürnberg, Erlangen, Germany

## ABSTRACT

The Cherenkov Telescope Array (CTA) is the next generation very high-energy gamma-ray observatory, with at least 10 times higher sensitivity than current instruments. CTA will comprise several tens of Imaging Atmospheric Cherenkov Telescopes (IACTs) operated in array-mode and divided into three size classes: large, medium and small telescopes. The total reflective surface could be up to 10,000 m$^2$ requiring unprecedented technological efforts. The properties of the reflector directly influence the telescope performance and thus constitute a fundamental ingredient to improve and maintain the sensitivity. The R&D status of lightweight, reliable and cost-effective mirror facets for the CTA telescope reflectors for the different classes of telescopes is reviewed in this paper.

**Keywords:** Cherenkov Telescopes, Cherenkov Telescope Array, mirrors, reflecting coatings

# 1. INTRODUCTION

With the advent of the Imaging Atmospheric Cherenkov Telescopes (IACTs) [1, 2] in late 1980's, ground-based observation of TeV gamma rays came into reality and, since the first source detected at TeV energies in 1989 the number of gamma-ray sources has rapidly grown up to 145 to date (see the catalogue at the link: http://tevcat.uchicago.edu/). All of them, a part a few cases, have been revealed at TeV energies with IACT telescopes.

The technique was first pioneered by the *Whipple* experiment since 1985, leading to the discovery of TeV gamma-rays from the Crab Nebula in 1989. This first result was followed by the discovery of the TeV emission from the first extragalactic source (Mrk 421), showing that acceleration processes are taking part in AGNs too. Since 2003, as the new generation experiments (HESS, MAGIC and VERITAS) have been started to observe the gamma-ray sky, the number of VHE sources started to rapidly increase. New classes of sources were detected at GeV-TeV energies both galactic (e.g. Galactic Center, Pulsar, Wind Nebulae, Pulsars and Binary Systems) and extragalactic (e.g. Blazars, radio-galaxies, star-forming galaxies) as well as about a dozen of unknown new TeV sources.

The recent advances of TeV γ-ray astronomy have shown that the 10 GeV – 100 TeV energy band is crucial to investigate the physics prevailing in extreme conditions found in remote cosmic objects as well as to testing fundamental physics. Nevertheless, after the launch of two gamma-ray dedicated satellites (AGILE and Fermi), the gamma-ray astronomy is now living a sort of *Golden Age* and opening unprecedented opportunities of multi-wavelength observations on a very wide energy range. In such an exciting scenario, a new generation of ground-based VHE gamma-ray instruments are needed in order to significantly improve the sensitivity, the operational bandwidth, the field of view and the angular resolution.

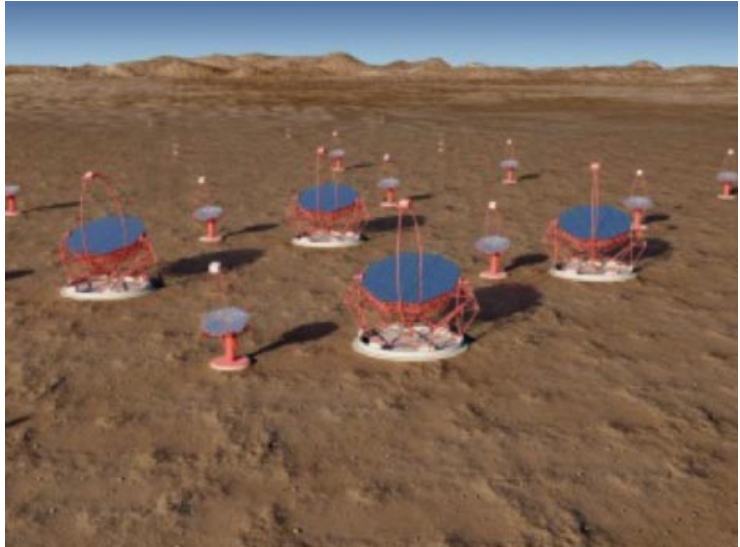

Fig. 1. Pictorial view of the CTA array at the southern site.

The international VHE astrophysics community is moving towards such a new generation of Cherenkov experiments and now both in Europe and USA the Agencies and a large consortium of Institutes support the implementation of the Cherenkov Telescope Array (CTA) observatory [3, 4]. CTA is conceived to allow both detection and in-depth study of large samples of known source types, and to explore a wide range of classes of suspected gamma-ray emitters beyond the sensitivity of current instruments. With its innovative approach based on the use of four different kinds of telescopes, CTA will obtain a one-order-of-magnitude improvement with respect to the current Cherenkov telescope performance. Moreover, CTA aims to provide global coverage of the sky from two observatory sites: a Southern array (see Fig. 1), implemented in particular for the exploration of both the Galactic plane and the extragalactic sky, and a Northern array, mainly devoted to the study of extragalactic sources.

In the atmospheric showers originated by a gamma–ray primary, the Cherenkov light intensity is almost proportional to the gamma–ray energy (see e.g. [5]). In general, large–diameter mirrors are needed to trigger on low energy gamma rays, while small mirrors are sufficient enough to trigger on high-energy events. Moreover, due to the very low gamma-ray fluxes at high energy, future Cherenkov telescopes like CTA must be able to catch events reaching the ground very far (∼ 300 − 500 m) from the telescope position, thus achieving effective areas of the order of $10^6$ m$^2$. In order to trigger far showers, imaged at large off-axis angles, Cherenkov telescopes must be provided with sufficiently large fields of view.

Optical dish diameters and field of view are the first parameters to be considered before analysing other specific aspects such as optical design, mirror structure, focal plane sensors and electronics. In this respect, due to the forward direction of Cherenkov light emission in air (and the resulting Cherenkov light pool of fairly uniform illumination of about 200 – 250 m diameter), for CTA an inter-telescope spacing of about 100 m is needed at threshold energies to provide images in multiple telescopes. Well above this threshold, showers can be detected from outside the light pool, if the field of view is large enough.

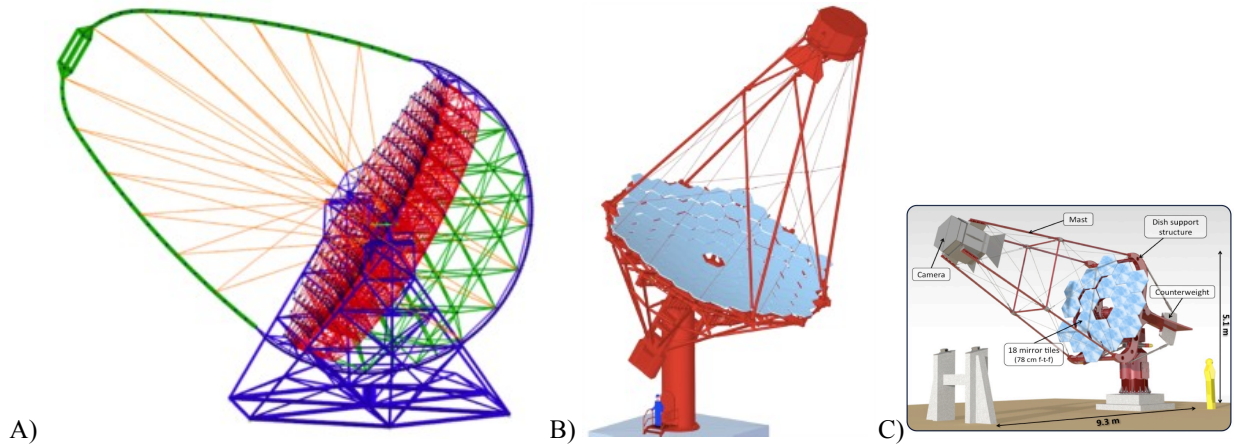

Fig. 2. Single-dish telescopes being developed for the CTA arrays. A) Large Size Telescope (LST); B) Medium Size Telescope (MST); C) Small Size Telescope single mirror (SST 1M).

In CTA a small number (4 units for both northern and southern sites) of Large Size Telescopes (LST) of 23 m diameter (see Fig. 2 - A) will be deployed close to the centre of both arrays, with ≃100 m spacing [6]. A larger number (up to 25 in the southern site, 15 in the northern site) of Medium Size Telescopes (MST, see Fig. 2 - B) will cover a larger area, with an inter-telescope spacing of ≃ 150 m [7]. Moreover the southern site will include up to 36 Schwarzschild-Couder dual-mirror Telescopes (SCT) of 9.5 m diameter [8, 9] (see Fig. 3 - A).

Above a few TeVs, the Cherenkov light intensity is such that showers can be detected even well outside the light pool by telescopes significantly smaller than the MST. To achieve the required sensitivity at high energies, a very large area on the ground needs to be covered by the Small Size Telescopes with a resolution of ≃ 0.1$^o$ and a field of view of ≃ 10$^o$. The SST sub–array can therefore be accomplished by 70 telescopes with a mirror area of 5–10 m$^2$ and ≃ 300 m spacing, distributed across an area of 10 km$^2$ and within a radius of about 3 km. The SST array will be implemented just on the southern site for reason of costs and taking into account that the very high energy emission can be observed just for galactic sources, unless non standard processes are invoked.

It should be noted that classical parabolic [10] or Davies-Cotton (DC) [11] single–mirror configurations have been used so far for Cherenkov telescopes, and they are adopted also in the case of CTA, for the LST and MST telescopes respectively. The mirror in both cases are made of tiles with spherical shapes (that, for the parabolic case, have different radii of curvature, while for D-C telescopes have all the same radius). However they are dominated by the cost of the camera, which is based on classical large-size photo-multipliers and they do not seem ideal for making the wide-field SST telescopes under development [12, 13] (see Fig. 2 - C). A possible solution, under study in the context of CTA, is to implement the classical DC solution also for the SST telescopes, with the addition of Winston-cone light-guides mounted within the camera to squeeze the light with an aggressive concentration ratio; however this solution could present some drawbacks, mainly regarding the difficult implementation and a limited number of pixels [14, 15]. Another attractive alternative solution to realize the SST telescopes is the use of a dual-mirror (2M) design, adopting the so-called Schwarzschild–Couder (SC) configuration [16]. This enables good angular resolution across the entire field of view, almost 10$^o$ in diameter, and also reduces the effective focal length and camera size. As has previously been demonstrated, 2M SC telescopes allow better correction of aberrations at large field angles and hence the construction of telescopes with a smaller focal ratio. The SST group within CTA is also developing 2M telescopes which have the potential to provide the required optical performance and allow exploitation of these technologies [17]. In this context two studies are in particular being carried out, one by the Italian ASTRI collaboration [18], based on a primary mirror made of hexagonal tiles and a monolithic secondary mirror of 2 m diameter (see Fig. 3 – B1), and the other by the GATE France-UK collaboration [19] based on a primary mirror made of petals and a segmented secondary mirror of 2 m diameter (see Fig. 3 – B2).

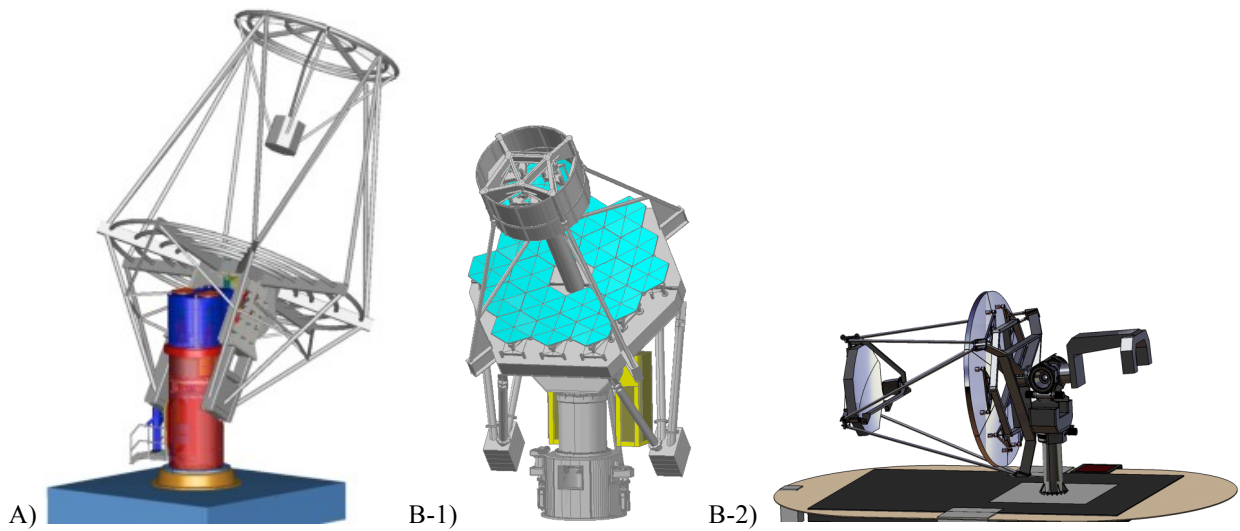

Fig. 3. Dual-mirror Schwarzschild- Couder telescopes being developed for the CTA arrays. A) Medium Size Schwarzschild -Couder Telescope (SCT); B-1) Dual Mirror Small Size Telescope (DM SST) – ASTRI design; B-2) Dual Mirror Small Size Telescope (DM SST) – GATE design.

In general, the telescopes of the two CTA arrays will have a combined mirror area of up to 10,000 m. The requirements for the focal point spread function (PSF) are more relaxed compared to those for optical telescopes. Typically, an overall PSF below a few arcmin is acceptable, which makes the use of a segmented reflector consisting of small individual mirror facets (called mirrors in the following) possible. It should be noted that the CTA telescopes will not be protected by domes and so the mirrors are permanently exposed to the environment. The design goal is to develop low-cost (aiming at < 2000 \$/m$^2$), light-weight, robust and reliable mirrors of 1-2 m size with adequate reflectance and focusing qualities but demanding very little maintenance.

Current IACTs mostly use polished glass (HESS [20] and VERITAS [21]) or diamond-milled Aluminum mirrors (MAGIC I and partially MAGICII [22, 23]), entailing high cost, considerable time and labor intensive machining. The technologies under investigation for CTA [24, 25] instead mainly pursue different methods such as sandwich concepts with cold-slumped surfaces made of thin float glass and different core materials like Aluminum honeycomb, glass foams or Aluminum foams, constructions based on carbon fiber/epoxy or glass fiber substrates, as well as sandwich structures made entirely of Aluminum.

For the telescopes with dual-reflector optics, SST-2M and SCT, as well as the single mirror reflector telescope and SST-1M, the primary and secondary mirrors have requirements which are more constraining than for the MST and LST designs. Short radii of curvature and strongly aspherical components are the main challenges to be faced. Also, for the primary mirror, the radius of curvature would be less than 10 m (down to a few meters for the secondary mirror), and the asphericity can reach deviations of tenths of a millimeter with respect to the best-fitting sphere for a single mirror segment. In this paper we will review the requirements for the production of the CTA mirrors and discuss technologies under development for the realization of the CTA mirrors. In this respect, we will treat just the efforts on-going for the production of single-mirror telescopes (LST, MST and SST-1M). Details about the realization of the reflectors for dual-mirror telescopes are reported on two other papers of these proceedings [26, 27]

## 2. SPECS, REQUIREMENTS & NEEDS FOR THE CTA MIRRORS

The main parameters, including those concerning the mirrors to be implemented, of the telescopes of the CTA array are reported in Tab. 1.

Table 1. Main parameters of the CTA telescopes under development.

| | SST 1M | SST 2M | MST | LST | SCT (MST SC) |
|---|---|---|---|---|---|
| **Number of units*** | 70 (S) | 70 (S) | 25 (S) 15 (N) | 4 (S) 4 (N) | 36 (S) |
| **Energy Range** | 100 TeV to 5 TeV | 100 TeV to 5 TeV | 200 GeV to 10 TeV | 20 GeV to 1 TeV | 200 GeV to 10 TeV |
| **Effective Area of mirrors per unit** | > 5 m$^2$ | > 5 m$^2$ | > 88 m$^2$ | 390 m$^2$ | > 40 m$^2$ |
| **Field of View** | > 8º | > 8º | > 7º | > 4.4º | > 7º |
| **Telescope Angular resolution (diameter 80 %)** | <0.25º | <0.25º | <0.18º | <11º | <0.075º |
| **Telescope Configuration** | Single Mirror Davies-Cotton | Dual Mirror Schwarzschild-Couder | Single Mirror Davies-Cotton | Single Mirror Parabolic (approximation) | Dual Mirror Schwarzschild-Couder |
| **Diameter of the primary mirror** | 4.0 m | 4.3 m (ASTRI) 4 m (GATE) | 12 | 23 m | 9.5 m |
| **Focal length** | 5.6 m | 2 m | 16 m | 31 m | 5.6 m |
| **Tiles of the primary mirror: number, configuration, & size** | 18, Hexagonal, 78 cm (flat-to-flat) | ASTRI: 18, Hexagonal, 85 cm (flat-to-flat) GATE: 6, Petals,1434 cm (max dimension) | 84, Hexagonal, 120 cm (flat-to-flat) | 207, Hexagonal, 151 cm (flat-to-flat) | 48, Petals, 125 cm to 145 cm (max dimension) |
| **Diameter of the secondary mirror** | Not Applicable | 1.8 m (ASTRI) 2 m (GATE) | Not Applicable | Not Applicable | 5.42 m |
| **Secondary mirror: configuration, number & size of tiles** | Not Applicable | ASTRI: Monolithic, 1, 1800 cm  GATE: Segmented, 9, 70 cm | Not Applicable | Not Applicable | Segmented, 24, 123 cm to 104 cm, |

*N.B.: S → CTA Southern site, N → CTA Northern site

IACTs are normally placed at altitudes of 1,000 − 3,000 m a.s.l., where significant temperature changes between day and night as well as rapid temperature drops are quite frequent. All optical properties should stay within specifications within the range −10◦ C to +30◦ C and the mirrors should survive temperature changes from −25°C to +60°C with all possible changes of their properties being reversible, as well as maximum wind speeds of 200 km/h. Intrinsic aberrations in the Cherenkov light emitted by atmospheric showers limit the angular resolution to around 30 arcsec. However, the final requirements for the resolution of the reflectors of future CTA telescopes, i.e. the spot size of the reflected light in the focal plane (camera), will depend on the pixel size of the camera and the final design of the telescope reflector. There is no real need to produce mirrors with a PSF well below 1/3 of the camera pixel size, which is ordinarily not smaller than 5 arcmin. A diffuse reflected component is not critical as long as it is spread out over a large solid angle. The reflectance into the focal spot should exceed 85% for all wavelengths in the range from 300 to 600 nm, ideally close to (or even above) 90%. The Cherenkov light intensity peaks between 300 and 450 nm, therefore the reflectance of the coating should be optimized for this range.

The mirrors for the CTA telescopes will be hexagonal in shape, with an anticipated size between 1−2 m, well beyond the common size of 0.3−1 m of the currently operational instruments. The weight should be ~20/40 kg depending on the telescope type.

### 3. MIRROR TECHNOLOGIES UNDER STUDY

Several institutes within the CTA consortium have developed or improved different technologies to build mirrors, most of which are in a pre-production phase at the moment. The technologies being studied are reported in Table 2.

Table 2. Technologies under study for the production of CTA mirrors.

| Mirror Technology | Description | Coating | Responsible Institution | Valid for: |
|---|---|---|---|---|
| Glass Replica Mirrors | Two glass sheets, cold slumped, with Al honeycomb in-between. | Al+SiO$_2$ (PVD) | INAF, Italy (collaboration with Media Lario Technologies company, Italy) | MST, MST SC primary |
| | Two glass sheets, cold slumped, two G10 sheets, with Al honeycomb in between, side walls | Al+Si0$_2$+HfO$_2$+SiO$_2$ (PVD) | IRFU, France (collaboration with Kendry company, France) | MST |
| | Two glass sheets, cold slumped, with Al honeycomb in between | Cr + Al + SiO$_2$+ HfO$_2$ + SiO$_2$ (sputtered) | ICRR, Japan (in collaboration with Sanko company, Japan) | LST, MST |
| | Two glass sheet (not bent) with Al spacers, open structure (not sealed), additional intermediate layers, additional cold slumped front glass sheet on epoxy layer | Al+SiO$_2$+HfO$_2$+SiO$_2$ (PVD) | IFJ-PAN, Poland | MST |
| | Two glass sheets, hot pre-slumped, cold slumped for fine adjustment of shape, with Al honeycomb in-between | Al+SiO$_2$ or multilayer SiO$_2$+TiO$_2$ (PVD) | INAF, Italy (collaboration with Media Lario Technologies company, Italy and Flabeg company, Germany) | SST -2M primary & secondary |
| | Hot-slumped monolithic glass | Al+SiO$_2$ or multilayer | INAF, Italy (collaboration with | SST -2M primary & |
| | | SiO$_2$+TiO$_2$ (PVD) | Flabeg company, Germany) | secondary |
| Ni electroforming and composite materials | TBD | TBD | Univ. of Alabama Huntsville, USA | SCT |
| Composite Mirrors | SMD technology (automotive industry) to form monolithic structure, gel-coat to form smooth surface | Al+SiO$_2$ (PVD) | SRC-PAS, Poland | SST -1M |
| Diamond Milled Aluminum Mirrors | Two Aluminum sheets, one diamond-milled to the desired profile | SiO$_2$ (PVD) on diamond milled Aluminum | INFN & Univ. of Padova | LST, MST |
| Polished Aluminum mirrors | Polished Aluminum mirrors with Ni/P Kanigen alloy cladding, layer | SiO$_2$ (PVD) | Observ. De Paris | SST -2M |
| Polished Glass Mirrors | Polished glass mirrors | Al+SiO$_2$ (PVD) | Joint Laboratory of Optics, Czech Republic | SST -1M |
| | Polished glass mirrors | Al+SiO$_2$ (PVD) | Galaktika company, Armenia | SST -1M, MST |
| | Polished glass mirrors | Al+SiO$_2$+HfO$_2$+SiO$_2$ (PVD) | Wilhelm Weule company, Germany | SST -1M |

## 4. MIRROR TECHNOLOGIES FOR TELESCOPES WITH SINGLE REFLECTOR OPTICS

### 4.1 Glass Replica mirrors

The cold glass slumping technique is a new method developed and set-up during the past few years by the Brera Astronomical Observatory of the Italian National Institute of Astrophysics (INAF-OAB) in collaboration with the Media Lario Technologies company [28, 29, 30]. This effort was used for the 17 m diameter MAGIC II telescope. A production rate of one panel per day per mould was successfully achieved (2 moulds were developed at this aim by INAF-OAB). The process exploits the concept of replication from a master to force the reflecting surface to assume of the desired shape. This is a very competitive approach for productions that aim at many identical pieces such as the case of mirrors for the Cherenkov telescopes with a Davies Cotton optical layout (this is the case of CTA for LST, MST and SST-1M). It should be noted that a similar technology was already proposed and studied for the production of mirrors for Cherenkov telescope in the nineties [31], while similar approaches are followed for making the reflectors for Cherenkov light in particle physics experiments (see e.g. reference [32]).

**The cold slumping approach followed at the INAF/Brera Astronomical Observatory (Italy)**

The approach has bee extensively described by Canestrari et al. [33, 34]. A thin glass sheet is bent by vacuum suction and is forced to adhere to a mould having a highly precise shape (the mould's profile is the negative of that desired on the mirror). The complete panel is assembled, by gluing a reinforcing core second thin glass sheet. Aluminum honeycombs are typically used for the core structure. After the glue is polymerized, the vacuum suction can be released and the substrate properly coated. The coating consists of two layers deposited by plasma vapour deposition - one of evaporated Aluminum and another of thin Quartz - as protection against reflectivity losses and scratches. The mirror is then finished by sealing its edges to prevent damage from water infiltration.

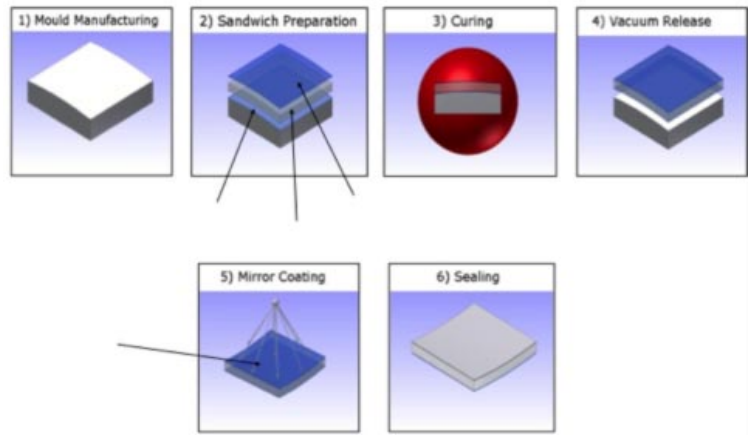

Fig. 4. Main production steps of INAF-OAB cold-slumped mirrors.

At the end of the process a mirror panel with a sandwich-like mechanical structure is obtained. It shows a high mechanical stiffness and low areal density (about 10 kg/m$^2$). Fig. 4 shows the main manufacturing steps. The development of the technology has been performed a few years ago for the MAGIC II experiment. However, a new investment has been done by INAF to get an overall adaptation to the requirements of CTA and new design for the MST. The realization of a pre-production series of about 25 mirrors is on-going in such a way that: 20 full specs mirrors have been delivered to the CTA consortium for installation and on-site testing at the MST prototype. Figure 4 shows the mirrors already produced and measured; 5 mirrors are being prototypes for new materials and/or coatings. The total amount of money invested also includes the complete requalification of the process and materials used by repeating all the qualification tests described in table. The technology has already been transferred to a company able to ensure the production needs of CTA.

**The Cold slumped mirror-sandwich design being pursued by IRFU/CEA, Saclay (France)**
The IRFU Saclay concept for CTA mirrors [35] is based on the replication of the shape of a mould, following a cold-slumped technique. The IRFU-Saclay mirrors consist of different layers of materials. A stiff core is sandwiched between two glass sheets, one of which is to receive the reflecting coating. The mirrors are front/back symmetrical in order to offer the best thermal behaviour. The core of the mirror is a 3-layer sandwich made of Aluminum honeycomb and two layers of glass fibre (G10). Glass fibre is chosen for its thermal dilation coefficient, which is very close to the one of the glass sheets. The layers are held together with glue and processed against the mould in order for it to have one spherically concave shaped face. The contact between the mould and the sandwich is obtained by vacuum suction.

As the Aluminum honeycomb is made of many cells, a specific micro-punched material is used for making the pressure equalized in each cell. This honeycomb is based on walls of thin 50 μm Aluminum foils. In that way it is rather flexible and does not need any milling to match the curvature of the mirror. As a second step, 2 mm thick glass sheets are glued on each face, still with the mirror held onto the mould.

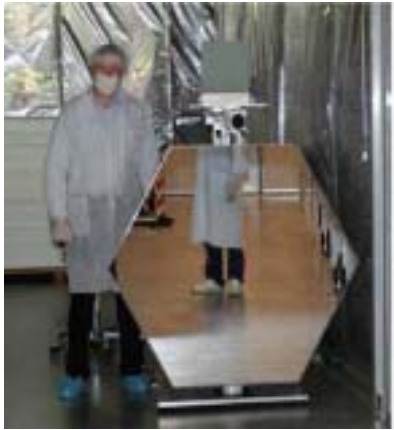

Fig. 5. MST mirror prototype produced by INAF in collaboration with Media Lario Technologies.

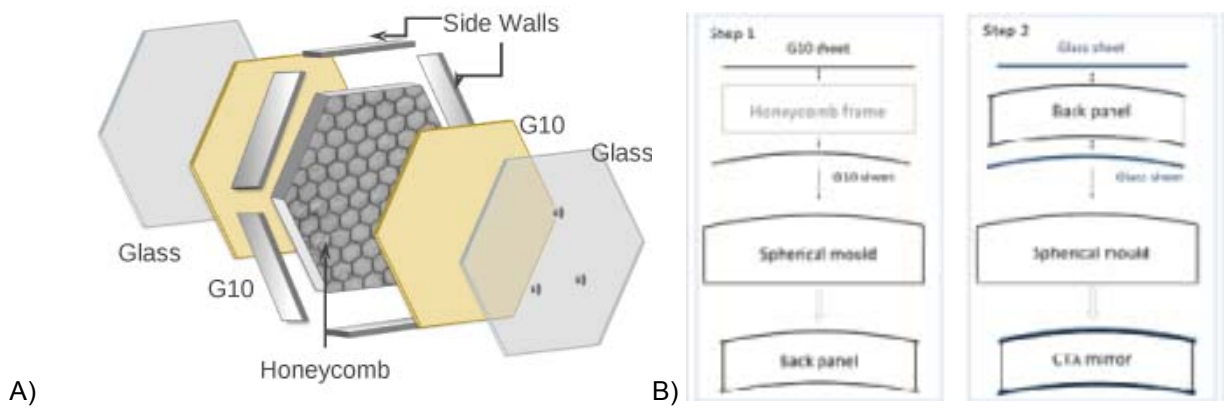

Fig. 6. A) Sketch of the mirror design and B) of the fabrication technique adopted by IRFU-Saclay.

During the assembly process thick sidewalls are placed on the six lateral sides; they will help to constrain the edges to bending when they are placed on the mould. The coating of the reflective face is deposited in collaboration with the Kerdry industry. Prototype mirrors have been already coated with Al+SiO2 or Al+Si02+HfO2+SiO2 films. Fig. 6 shows the sketch of the mirror design (A) and of the fabrication technique (B).

About 30 prototype mirrors of hexagonal shape, 1.2 m face to face have been produced with this approach, to develop and refine the technology. Those mirrors are spherical with 33.6 m radius of curvature; they weigh less than 25 kg and have a thickness of 85 mm at most. The prototype mirrors have been tested optically. They fulfill the specifications for CTA MST mirrors. The focal lengths of the mirrors as measured on the optical test bench correspond to half the radius of the mould as expected, with an accuracy of a few cm. The PSF and reflectivity are within the CTA specifications. In Figure 7 a picture of one of the mirrors is reported (20 MST mirror prototypes have been produced)

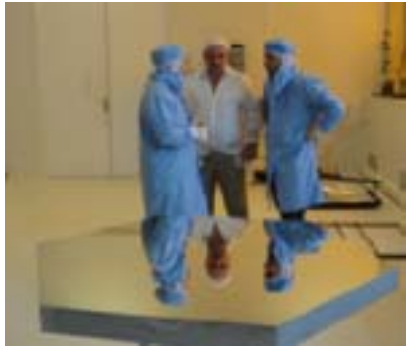

Fig. 7. One of the IRFU-Saclay mirrors produced for the MST.

**The cold slumping approach followed at ICRR, Japan (in collaboration with the Sanko industry)**

Large-size tiles have to be produced for the LST and they are being developed in Japan at ICRR (Japan) in collaboration with the Sanko industry. They are pursuing a cold slump technique, pretty similar to the approach used by INAF for the production of the glass mirrors of the MAGIC-II telescope (see Fig. 8-A). Prototypes have been already successfully produced (see Fig. 8-B). The mirror facets have a sandwich structure consisting of a glass sheet, an Aluminum honeycomb and another glass sheet. The main production steps are described hereafter:

1. The rear glass sheet of 2.7 mm thickness is placed on the mould, which has the curvature of about 56 m. The glass sheet is continuously sucked by vacuum to the mould to keep the curved shape during all process.

2. The SUS410S pre-formed hexacell shape frame is glued on the glass sheet to define the hex cell shape "mirror box". The SUS410S frame is pre-machined to fit to the curve of mirrors.

3. The 60 mm Aluminum honeycomb is filled inside this hex box, and glued together with rear glass sheet. The hex-shape "mirror box" is sealed with this rear glass.

4. Pressure of 500kg/m$^2$ is applied on the rear glass sheet with a dummy plate.

5. For the curing of the glue, the mirror structure and mould are kept in the temperature box of 80°C for two hours. Aluminum honeycomb with slits is used, which allow the water and air can move freely between hex-cells. Two drain holes are prepared at the bottom of the mirrors.

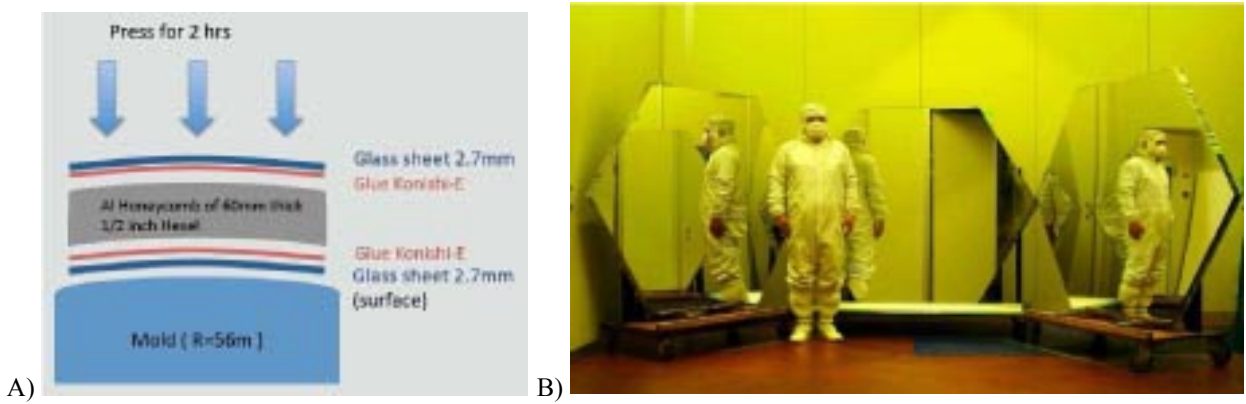

Fig. 8. A) The segmented mirrors with sandwich structure are produced with the cold slump technique (replica method). B): The first 1.5m flat-to-flat hexagonal shape mirror produced at the company Sanko in Japan.

**The open-structure composite mirror design from the Institute of Nuclear Physics of the Polish Academy of Sciences (IFJ PAN, Poland)**

The open-structure mirror [36] consists of a rigid flat sandwich support structure, aspherical resin layer and a spherical glass reflecting layer, as shown in Figure 9-A. The sandwich support structure is built of two flat glass panels interspaced by v-shaped Aluminum spacers of equal length. They are glued to the panels with the epoxy resin. A compensation glass sheet with the resin layer is glued to the rear panel to ensure the geometrical stability and stiffness of the structure. Also, two fibre-glass tissues are used to reinforce the substrate structure and to improve the resistance to the mechanical impact. One of the fibre-glass tissues is placed in between the spherical reflective layer and the spherical epoxy layer. Another tissue is in between the flat layer of the epoxy resin and the glass compensation layer. Finally, three stainless steel pads are glued to the compensation glass layer as an interface for mounting of the actuators designed for CTA mirrors. The sandwich support structure thus built is an open structure and as such it represents a novel approach, different from the commonly applied closed Aluminum honeycomb support. The latter requires a perfect sealing of the mirror so that water cannot penetrate inside the honeycomb and possibly cause damage to the structure. The open sandwich structure with open-profile spacers enables good cooling and ventilation of the mirror panels and also prohibits water to be trapped inside - the spacers are organized such that the water flows almost freely through the structure. To protect the structure against contamination by an insect or bird wasteand also against hard rime, a stainless steel mesh is attached to the side walls. To obtain the convex surface of the mirror a layer made of compound of the epoxy resin and fillers is cast onto a front panel in a high-precision mould which is specifically designed for this purpose. After hardening at room temperature, the resin layer reproduces the exact shape of the mould. Then a reflective layer made of Borofloat 33 glass sheet is cold-slumped to the spherical resin layer in the same mould. The Borofloat 33 glass reflective layer is coated prior to gluing to the substrate. In the last step a silicone rubber is glued to the mirror sidewalls to protect the mirror against damage during the transportation and mounting on the telescope structures. The silicone rubber serves also as a protection against the water penetration into the epoxy resin layers. Note, that the technology developed at IFJ PAN allows for the use of a dielectric coating, which provides high durability of the reflective surface. So far eleven mirrors aim for the MST prototype and for testing purposes have been produced this way, with ten of them having an $Al+SiO_2+HfO_2+SiO_2$ coating and one a dielectric coating (Fig. 9-B).

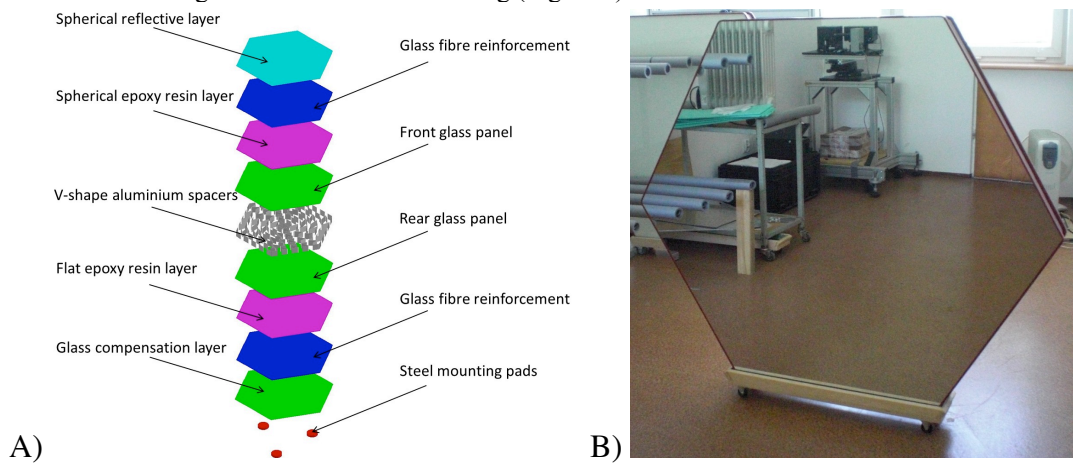

Fig. 9. A) Exploded view of the composite mirror design investigated by IFJ PAN. B) Picture of a composite mirror prototype developed by IFJ PAN.

**4.2 Polished Glass Mirrors**

**Moulded glass mirror design from Joint Laboratory of Optics, Olomouc (Czech Republic)**

The production technology of mirror segments in the Olomouc laboratory starts from a moulded glass semi-finished product made of SIMAX glass (a material with properties similar to PYREX). The semi-finished blanks are purchased from the KAVALIERGLASS glass factory, in Czech Republic. Afterwards, all the operations performed on the semi-products (including also the reflecting coating deposition) are carried out at the Joint Laboratory of Optics operated by the Palacky University and Institute Physics of the Academy of Sciences of the Czech Republic. The already existing classical optical polishing technology will be used for the scope. An improved technology based on CNC polishing is also being implemented. The production process foresees the cutting, milling, grinding and polishing operations for each mirror tile; all segments are thoroughly tested after each technological step. It should be noted that using this approach mirror segments of different sizes and radii of curvature have been already developed for several previous astroparticle experiments (e.g. CAT, CELESTE, AUGER, AUGER-HEAT) and particle physics projects (DIRAC), with typical production rate of one segment per day.

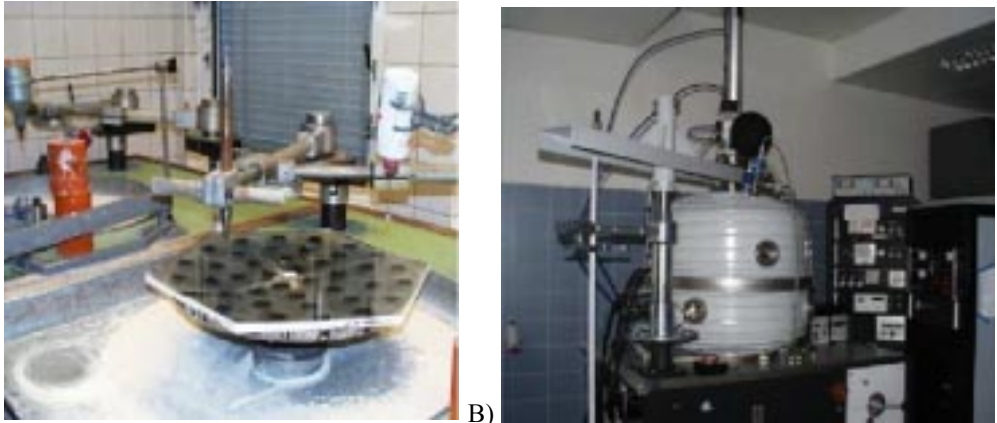

A)  B)

Fig. 10. The existing polishing machine in Olomouc lab and apparatus to deposit reflective and protective layer

The polishing machine in the laboratory is shown in Fig. 10-A. The technology to be used for CTA is based on an ultra-precise polishing with a CNC machine equipped with a bonnet tool with a variable internal pressure. The control of the shape parameters will be available using a sub-aperture interferometer that will provide immediate feedback on the achieved profile during the polishing process. The existing technology can produce mirror segments up to 25,000 mm of radius of curvature without any major additional problem. The diameter of circular segments or diameter of circumscribed circle to hexagonal segments varied from 500 mm to 623 mm. Segments with thickness from 6 to 16 mm were produced. The CNC polishing technology will be able to machine segments up to 1,000 mm in diameter for spherical surfaces and up-to 600 mm for aspherical ones. The limitation of radius of curvature is determined only by the thickness and size of the segments. The segment thickness needed for the CNC technology ranges between 5 and 20 mm. The vacuum deposition of optical thin layers is also available within the same lab (Fig. 10-B).

**Armenian Mirrors from Galaktika, Armenia**

The glass technology of polished moulded glass mirrors is well known and it has been used for the H.E.S.S Telescopes. The Galaktika CJSC company seems to be able to deliver mirrors which will meet CTA mirror requirements. The timeline of the mass production should be carefully defined with the company since glass mirror technology is time consuming.

**4.3 Composite mirror: the Sheet Moulding Compound mirror design from the Space Research Center (SRC) of the Polish academy of Science (SRC-PAS, Warsaw, Poland)**

Sheet Moulding Compound (SMC) is a low-cost, widespread and semi-fabricated product used for compression moulding being developed by Space Research Center (SRC), Warsaw, Poland (Fig. 11). The idea is to form mirrors for CTA telescopes during such a SMC compression moulding process. It is a technology derived by applications in the automotive industry. The main advantage is related to the quick production process, which would take just a few minutes, in principle profitable up to tens of thousands of pieces. In the process the top surface of the composite does not require polishing, since its smooth surface is obtained through high gloss mould. Therefore, the most difficult task in this technology is to get an optical surface directly on the SMC composite.

The SMC technology allows for a variety of solutions; therefore, three forming processes are being investigated at the moment:
1) Formation of the mirror at high pressure (60 bar) and high temperature (140-160° C) as a uniform ribbed

structure made of SMC (CF reinforced) for a base and with the top In-Mould Coating (IMC) layer for further reflective layer application;
2) Sandwich structure with Aluminum honeycomb as a core and SMC (CF reinforced) for the outer layers. Additional foil (Alanod) is used for reflective layer. The technology allows for low pressure (6 bar) and high temperature mould forming;
3) Reinforced structure with honeycomb or ribs as a core and SMC for the outer layers. Additional intermediate layer is used as a base for reflective coating. The technology allows for low pressure (6 bars) and low temperature mould forming.

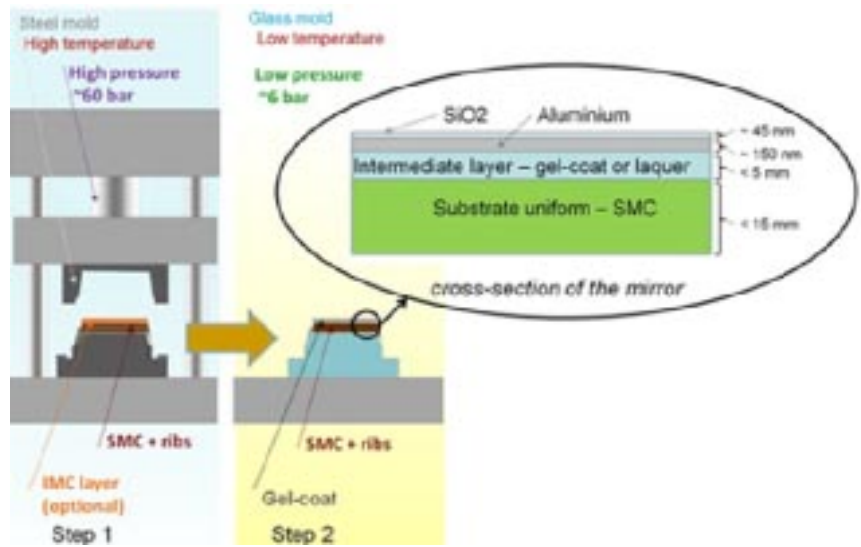

Fig. 11. Forming of composite mirrors - SMC technology with steel mould (step 1 - left) and application of an intermediate layer (step 2 - right) with indicated cross-subsection of layers (in the ellipse). The structure of the mirror facet is shown in the oval insert.

The prototyping phase is on-going at the moment (Fig. 12); all variants are investigated in parallel. The ribbed SMC structure requires development of a special mould and support from industry. Moreover, three sandwich structures (variant 3; scale 1:1) have been manufactured. Nevertheless, the intermediate and reflective layer developments need some further investigation. Since no glass is used as a front surface of the substrate, the coating includes not only reflective and protective layer but also an intermediate layer applied in order to reduce surface roughness of the composite substrate, improve mechanical resistance, and coating adhesion.

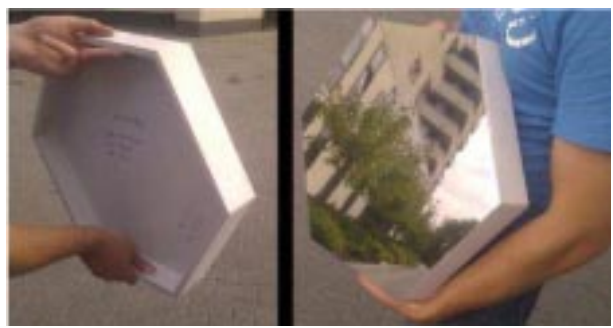

Fig. 12. The SMC mirror prototype developed by Space Research Centre, Warsaw.

**4.4 All-Aluminum sandwich mirrors developed by INAFN and Univ. of Padova (Italy)**
The entire reflector of MAGIC I and more than half of the MAGIC II mirrors are made of a sandwich of two thin

Aluminum layers interspaced by an Aluminum honeycomb structure that ensures rigidity, high temperature conductivity and low weight [23] (Fig. 13). The assembly is then sandwiched between spherical moulds and put in an autoclave, where a cycle of high temperature and pressure cures the structural glue. The reflective surface is then generated by precision diamond milling. A sketch is shown in Fig. 12 The final roughness of the surface is around 4 nm and the average reflectance is 85%. The Aluminum surface is protected by a thin layer of quartz (with some admixture of carbon) of around 100 nm thickness. For CTA, this technology is being further developed, especially by the use of either a thin coated glass sheet as the front layer or a reflective foil to reduce the cost imposed by the diamond milling of the front surface.

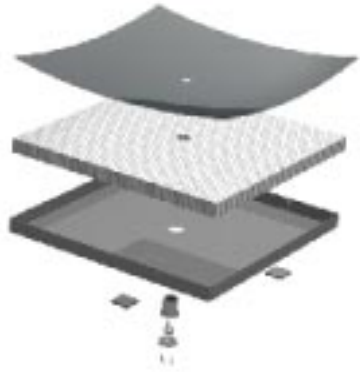

Fig. 13. Aluminum sandwich mirror being developed at INFN and Univ. of Padova.

## 5. COATINGS

IACTs need to have a good reflectance between 300 and 600 nm wavelengths, which makes Aluminum the natural choice as reflective material. The mirrors are exposed to the environment all year round, therefore this Aluminum coating is usually protected by vacuum deposited $SiO_2$ (in the case of H.E.S.S.), $SiO_2$ with carbon admixtures (for MAGIC) or $Al_2O_3$ obtained by anodizing the reflective Al layer (in the case of VERITAS). Nevertheless, a slow but constant degradation of the reflectance is observed. Not being protected by a dome, the mirrors are constantly exposed to the environment and show a loss of reflectance of a few percent per year. This requires re-coating of all mirrors every ~5 years. For the future CTA 2 observatory with~10,000 $m^2$ mirrors this would mean a significant maintenance effort.

To enhance the reflectance and the durability of the coatings two commercially available options are under investigation for MST [38]:

a) A three-layer protective coating ($SiO_2$ + $HfO_2$ + $SiO_2$) on top of a classic Al coating. This enhances the reflectance by ~5%.
b) A dielectric coating, consisting of a stack of many alternating layers of two materials with different refractive indices, without any metallic layer. This allows to custom-tailor a box-shaped reflectance curve with >95% reflectance between 300 and 600nm, and <30% elsewhere.

The design could be adjusted to the required wavelength range such, that e.g. a cut-off at 550nm allows us to reduce the night-sky background (first emission line around 556 nm). Fig. 14 shows the specular reflectance of these coatings compared to a classical Al+$SiO_2$ coating.

The coatings have been exposed to cycles in temperature (-10°C < T < 60°C; 5 h cycle duration) and humidity (5% to 95%; 8 h cycle duration) for a total of ~8000 h. The samples have been exposed for 72h to a salt-fog atmosphere at a temperature of ~20°C with a concentration of 5%. Again the samples with the simple $SiO_2$ protection show the strongest loss. A standard cheesecloth test using a force of 10N and 50 strokes shows a significant amount of scratches on the SiO2 sample, up to half of the SiO2 layer thickness in depth, while the two other coatings are hardly affected. For a more severe eraser test all samples show scratches, the SiO2 samples more than the three-layer samples and those more than the dielectric samples. The same order of resistance is revealed by a sand-blasting test.

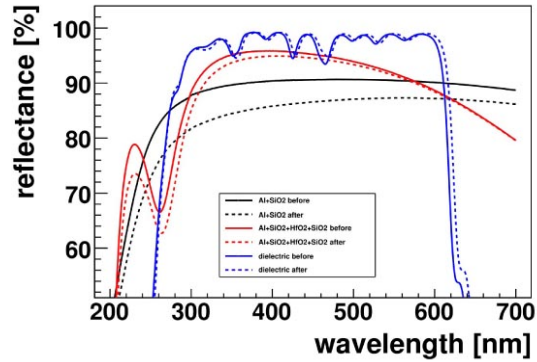

Fig. 14. Specular reflectance of the different coatings studied for MST.

In conclusion, both new options seem to be more resistant to environmental impact than the classical $SiO_2$ coating. The three-layer coating is being used in the current re-coating of the mirrors of the H.E.S.S. experiment (~1600 mirrors in total) with 100 mirrors getting the dielectric coating. Thus, real outdoor experience will be available soon, that will also allow us to verify possible problems of dew formation on the dielectric films, due to their high emissivity in the infra-red [39].

It should be noted that the investigation for the use of dielectric multilayer coatings is also being pursued for dual-mirror telescopes. In particular, for the ASTRI SST prototype fully dielectric multi-layer coatings have been developed and tested as an option for the primary mirror, aiming to filter out the large Night Sky Background contamination at wavelengths $\lambda > 700$ nm [40, 41], see Fig. 15.

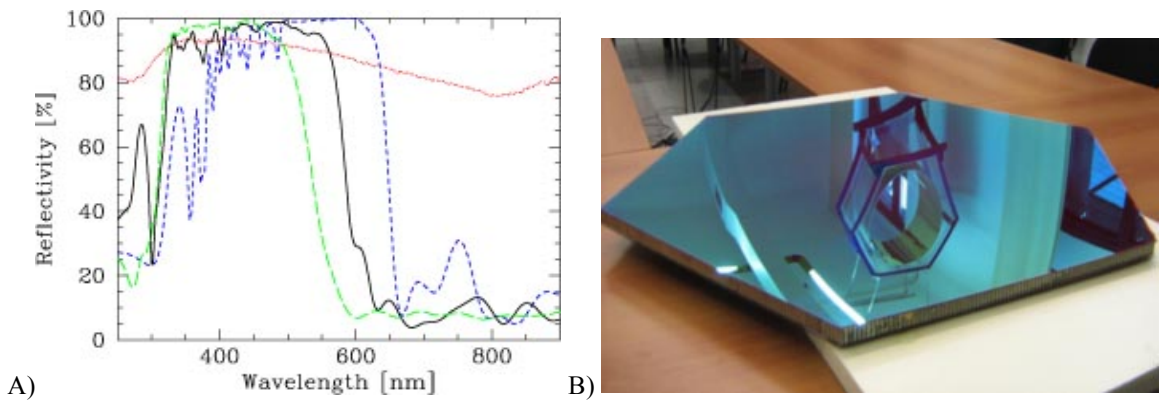

Fig. 15. Coatings under development for the ASTRI primary mirror. A) Measured reflectivities for the ASTRI primary mirror under normal incidence for the considered coating designs: Al+$SiO_2$ (dotted red), $SiO_2$+$TiO_2$ multilayer (short dashed blue), and $SiO_2$+mixed multilayer "orange" (solid black) and "yellow" (long dashed green). B) prototype segment of the primary mirror for the ASTRI telescope coated with $SiO_2$+$TiO_2$ multilayer.

## 6. OPTICAL MEASUREMENTS

The sensitivity of a Cherenkov telescope is closely related to the optical performance of its reflector. The figure of merit of the facet is characterized by the point spread function (PSF) and the absolute reflectance, depending on the wavelength in the 300 - 600 nm range, where most of Cherenkov light from atmospheric showers is concentrated.

- **PSF:** The detailed shape of the PSF is not of critical importance for Cherenkov telescopes as long as it is small enough. Therefore, the PSF is sufficiently characterized by the definition of the diameter of the circle that contains 80% of the reflected light ($d_{80}$). Since it is required that the $d_{80}$ should to be smaller than 1/3 of the camera pixel size, if for MST we assume a pixel size of 50 mm the $d_{80}$ has to be better than 17 mm (3.6 arcmin).

- **Absolute reflectance**: The absolute reflectance is defined as the fraction of the light incident onto the mirror surface that is reflected into the focal spot (i.e. it does not include the light scattered at large angles from the optical axis). There are different types of methods to measure PSF and absolute reflectance. One is the so-called "2f- method" currently employed with different levels of complexity and accuracy for resolution and/or reflectance measurements by most groups designing mirror substrates. A sketch for such a setup can be seen in Fig. 16.

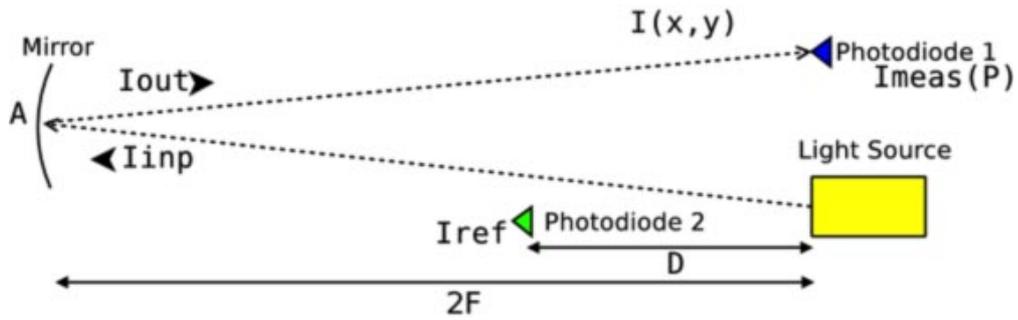

Fig. 16. Sketch of the 2-f measurement method.

A point-like light source is placed at twice the focal distance (i.e. the radius of curvature, for spherical mirrors) away from the mirror. At the same distance a detector system, usually a photodiode or CCD is used, is placed to record the return image. From the analysis of the recorded image, the PSF is determined. To calculate the absolute reflectance into the focal spot, the determination of the amount of incident light is additionally required. Several institutes within WP MIR have the facilities to perform 2-f measurements, performed for full size prototypes produced by INAF-Brera, CEA –Saclay, ICRR, Japan, and IFJ PAN. The measurement performed so far have shown very promising results, with PSF within the specifications.

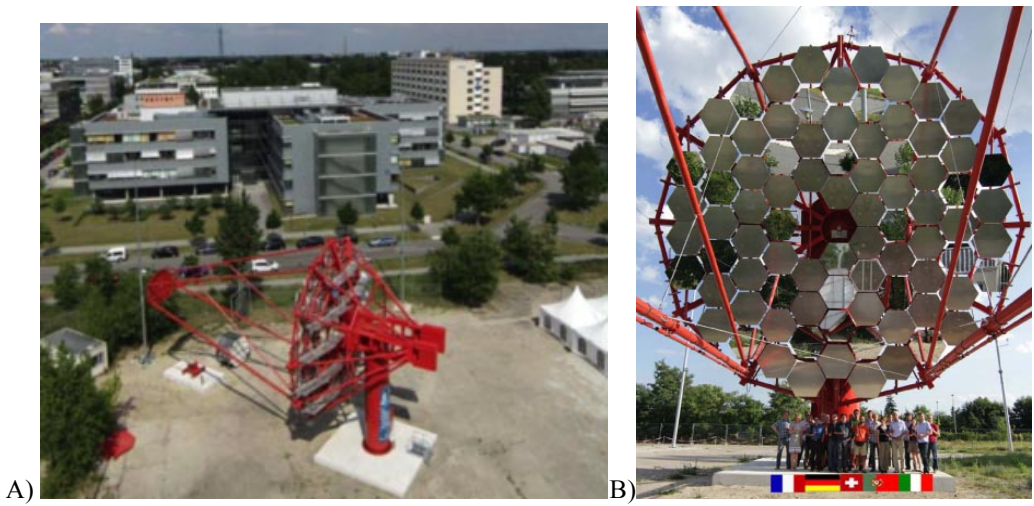

Fig. 17. The MST prototype installed in the DESY campus (Berlin). A) An aerial view of the telescope. B) the primary mirrors where part of the faced are dummy reflectors and part fully representative mirror tiles.

# 7. CONCLUSIONS

Many of the challenges induced by the demand for a few thousand mirrors with a total reflective area of up to 10,000 m$^2$ for CTA have been technically solved on the prototyping level, such as the production of large size facets of up to 2 m$^2$ in area with low weight and high optical quality. Currently pre-production series have been or are being produced for the different technologies to prove that an easy and rapid series production at reasonable costs is possible. For the mirror coatings, alternatives to the standard solution have been developed that show improved durability in laboratory tests. For quality control, extensive test facilities have been set up, as well for testing the optical performance of the mirrors as for determining the durability of substrates and coatings that will be used in a harsh outdoor environment for many years. A trade off among the different technologies will be performed soon after that all the durability and optical tests will be completed.

It should be noted that a full-scale prototype of the MST structure including drive and safety system was realized in the last two years; it is set-up in Berlin, in the campus of the DESY Institute (see Fig. 17). The primary mirror was also implemented, formed in part of dummy reflectors and part of fully representative reflecting tiles (produced by INAF-Brera, CEA Saclay mirrors, and IFJ Cracow). The measurements, surveys, and tests of the telescopes subsystems, including mirrors, are on-going.

## AKNOWLEDGMENTS

We gratefully acknowledge support from the agencies and organizations listed in this page: http://www.cta-observatory.org.